# Design and Status of JUNO

**Hans Theodor Josef Steiger** *on behalf of the JUNO Collaboration*

Physik-Department, Technische Universität München, James-Franck-Str. 1, 85748 Garching, Germany

hans.steiger@tum.de

**Abstract**. The Jiangmen Underground Neutrino Observatory (JUNO) is a 20 kton multi-purpose liquid scintillator detector currently being built in a dedicated underground laboratory in Jiangmen (PR China). JUNO' s main physics goal is to determine the neutrino mass ordering using electron anti-neutrinos from two nuclear power plants at a baseline of about 53 km. JUNO aims for an unprecedented energy resolution of 3% at 1 MeV for the central detector, with which the mass ordering can be measured with 3 – 4 σ significance within six years of operation. Most neutrino oscillation parameters in the solar and atmospheric sectors can also be measured with an accuracy of 1% or better. Furthermore, being the largest liquid scintillator detector of its kind, JUNO will monitor the neutrino sky continuously for contributing to neutrino and multi-messenger astronomy. JUNO's design as well as the status of its construction will be presented, together with a short excursion into its rich R&D program.

## 1. The JUNO Project – An Overview

The Jiangmen Underground Neutrino Observatory (JUNO) is a Liquid Scintillator Antineutrino Detector currently under construction within a dedicated underground laboratory (~700 m deep) close to Jiangmen city (Guangdong province, PR China). After the completion, it will be the largest liquid scintillator detector ever built, consisting 20 kt target mass made of Linear Alkyl-Benzene (LAB) liquid scintillator (LS), monitored by about 18000 twenty-inch high-quantum efficiency (QE) photo-multiplier tubes (PMTs) and about 26000 three-inch PMTs providing a total photo coverage of ~78%. Large photo coverage and QE in combination with a high light yield (~$10^4$ Photons/MeV) and excellent transparency (attenuation length > 20 m at 430 nm) of the LS are key parameters for the experiment. An unprecedented energy resolution of 3% at 1 MeV is in range. According to the conceptual design report [1] the LS will be stored in a 12 cm thick highly transparent acrylic hollow sphere with a diameter of 35.4 m. This construction is immersed in a ~44 m deep pool containing ~35 kt of ultrapure water (see Fig. 1). The water volume is instrumented with about 2000 additional twenty-inch PMTs acting as a Cherenkov detector to tag and veto cosmic muons.

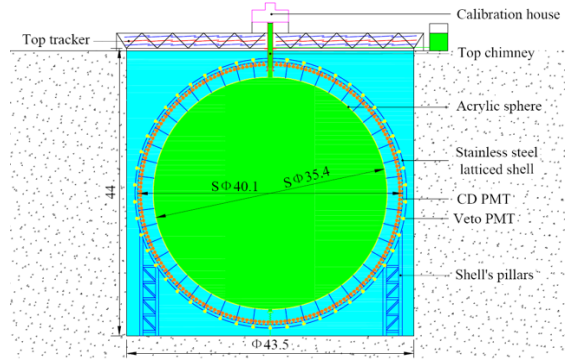

*Figure 1: Scheme of the JUNO central detector. A description can be found in the text.*

JUNO's main physics goal is the study of neutrino oscillations by detecting the disappearance of reactor $\bar{\nu}_e$ emitted by two nuclear power plants (Taishan & Yangjiang), both in 53 km distance to the detector. The nominal power of 26.6 GW$_{th}$ will be available from 2020 on, while a final thermal power of 35.8 GW$_{th}$ will be realized in the following years. The neutrino mass ordering (MO) can be measured with 3 – 4 σ significance within six years of operation. Most neutrino oscillation parameters in the solar and atmospheric sectors will also be measured with an accuracy of 1% or better. Furthermore, JUNO will measure Geo-neutrinos, contribute to the search for the DSNB (Diffuse Supernova Neutrino Background) and it will monitor the neutrino sky continuously for contributing to neutrino and multi-messenger astronomy. Beside this, JUNO will perform a proton decay search in the channel $p \rightarrow K^+ \bar{\nu}$. A complete description of JUNO's physics program can be found in [2].

## 2. Selected Aspects of the Detector Design

Within this section the requirements for a successful determination of the neutrino mass ordering and operation of JUNO as multi-purpose neutrino observatory will be discussed as well as some selected aspects of the detector design, that are foreseen to meet these requirements, will be illustrated further.

*2.1 Reactor Neutrino Detection and Requirements to the Detector Design*
The signature of the electron anti-neutrinos emitted by the reactors will be the inverse beta decay (IBD) process $\bar{\nu}_e + p \rightarrow e^+ + n$ in the JUNO detector, where p is a proton from the LS. The visible signal consists of two components, one prompt - the positron energy loss and annihilation and one delayed, when the neutron is captured, emitting a fixed energy of 2.2 MeV.
The resolution of the positron energy is one of the crucial parameters for the experiment, since the fine structure of the neutrino oscillations (fast component driven by $\Delta m_{31}^2$ as well as $\Delta m_{32}^2$ and slow component originating from $\Delta m_{21}^2$ at the medium baseline realized in JUNO) can only be resolved with an energy resolution better than $\Delta m_{21}^2 / |\Delta m_{31}^2|$. Furthermore, JUNO's sensitivity to the mass ordering is highly affected by the resolution as well as the event statistics (luminosity) [2].

*2.2 The PMT Arrays*
The scintillation light is red out in JUNO via two independent PMT arrays, one consisting of about 18000 twenty-inch high-quantum efficiency PMTs and a second independent one with almost 26000 small three-inch PMTs. Their arrangement can be seen in Fig. 2.

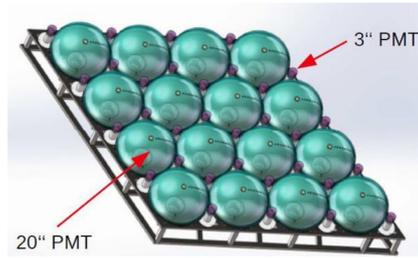

*Figure 2: Arrangement of the PMTs on one of the holding scaffolding panels. The 20-inch PMTs are drawn in green. The small 3-inch PMTs (in purple) will be mounted in the in the gaps of the laying pattern of the large tubes to optimize the photo-coverage to an absolute value of ~78%.*

The large PMT array contains two 20-inch PMT types; the Hamamatsu R12860 HQE optimized for high quantum efficiency (QE) and tubes using the multi-channel plate technology manufactured by Northern Night Vision Technology (NNVT). The small PMT array will be used as a second handle on the energy determination (double calorimetry) of an event. This will be beneficial for cross-calibration of the large PMT system, will reduce nonlinearities and will give the detector a greater dynamic range and granularity. For example, muon tracking and shower calorimetry, as well as supernova readout and solar oscillation parameter measurements are supposed to gain from the small PMTs. In the array a 3-inch PMT custom design (XP72B22) provided by HZC-Photonics (Hainan, PR China) with 24 % QE, an excellent single photon peak to valley ratio of 3.0:1 and a transition time spread of 2-5 ns will be used. The earth magnetic field will be compensated by dedicated coils in the detector.

### 2.3 PMT Testing

While the small PMTs are tested by HZC in corporation with the JUNO Collaboration at the factory site, the large PMTs are stored and tested in Zhongshan Pan-Asia. There dedicated performance testing systems were installed. Two container test systems with electromagnetic shielding are currently used for mass acceptance tests. Each of them consists of 36 individual drawers (see Fig. 3 a and b). The PMTs can be pulsed with a self-stabilized LED or a picosecond laser, both with 420 nm wavelength. A commercial switched-capacitor ADC (CAEN V1742) is used for waveform readout. Triggering and noise counting is realized via VME based leading edge discriminators (CAEN V895B) and latching scalers (CAEN V895AC). Using a PCIe computer bridge card and a custom made LabView based software, the entire system is controlled.

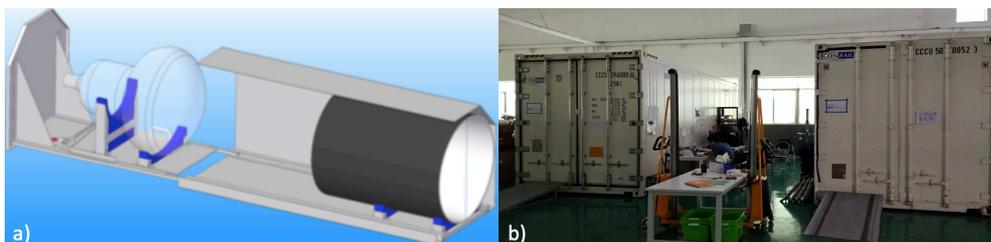

*Figure 3: a) One of the PMT drawers as CAD Model. b) The two PMT testing containers in Zhongshan Pan-Asia.*

In the containers the parameters charge resolution, single PE peak/valley ratio, operating voltage necessary for a gain of $10^7$, dark count rate (DCR), single PE rise and fall time as well as pre pulsing and after pulsing rate are determined. The PDE of the PMTs already tested up to July 2019 (5000 Hamamatsu and ∼7500 NNVT) are shown in Fig. 4 (the other parameters can be found in [3]).

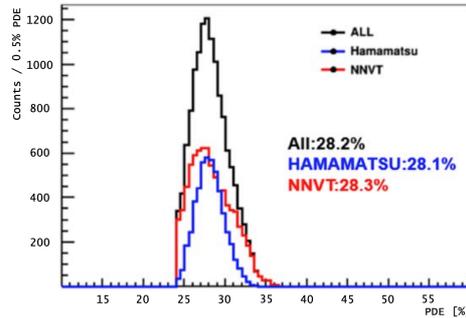

*Figure 4: PDE of 5000 Hamamatsu (blue) and ~7500 NNVT (red) 20-inch PMTs tested in the Container system. The black curve represents the sum over the histograms for both PMT types.*

Beside the containers, two scanning stations were set up to irradiate the photo cathodes from 168 point-like sources. The scanning provides information about the non-uniformity of PMT parameters, as well as their earth magnetic field dependence (for a detailed description see [4]).

*2.4. Liquid Scintillator*

The LS mixture consists of three components: LAB as a solvent, 2.5 g/l PPO (2,5-Diphenyloxazole) acting as fluor and 1-3 mg/l bis-MSB (bis-MSB [1,4-bis-(o-methylstyryl)-benzene]) as secondary wavelength shifter. The optical requirements are a light yield of $\sim 10^4$ photons per MeV energy deposition and an attenuation length for 430 nm light above 20 m. For reactor neutrinos, mass concentrations of $^{238}$U and $^{232}$Th below $10^{-15}$ g/g and for $^{40}$K $10^{-16}$ g/g are required. For effective solar neutrino spectroscopy these concentrations should be two orders of magnitude lower, while the amount of $^{14}$C should not exceed $10^{-18}$ g/g. To achieve these values, the LS will be purified in a series of dedicated process. In a first step the LAB will be filtered in an $Al_2O_3$ column. After that, it will be distilled to further improve the transparency and remove heavy radioactive metals. In the following $^{40}$K and more isotopes from the uranium and thorium chain will be removed by water extraction. Gaseous impurities like argon, krypton and radon are removed by steam stripping. A detailed description of distillation and stripping systems and first results from a pilot plant test phase at the Daya Bay detector site can be found in [5]. The Online Scintillator Internal Radioactivity Investigation System (OSIRIS) is being developed as a failsafe monitor to aid the commissioning of the on-site scintillator purification plants and assess the quality of the scintillator batches in terms of radioactivity before filling them into the central detector of JUNO. OSIRIS will be a liquid scintillator detector (with 19 t LS target) determining uranium and thorium concentrations via the fast Bi/Po-coincidence signals of this chains. Other contaminations (e.g. $^{14}$C, $^{85}$K) will also be evaluated. According to Monte Carlo studies radioactivity measurements down to the level necessary for the JUNO IBD requirements will be possible within 10 hours, while the solar limits could be reached within 5 days.

**References**
[1]  F. An et al. [JUNO Collaboration], *JUNO Conceptual Design Report*. [*arXiv:* 1508.07166]
[2]  F. An et al., *Neutrino Physics with JUNO*. J. Phys. G 43 (2016), 030401
[3]  H. Zhang et al., *Tested Performance of JUNO 20'' PMTs*. TAUP2019, Contribution ID: 103
[4]  Anfimov N., *Large photocathode 20-inch PMT testing methods for the JUNO experiment*. JINST, 2017, 12(06): C06017.
[5]  P. Lombardi et. al., *Distillation and stripping pilot plants for the JUNO neutrino detector:Design, operations and reliability*. NIM, A, 925 (2019), 6-17, [*arXiv:* 1902.05288]